# Unified Lepto-Quark Mixing

Alexander N. Jourjine[1]

**Abstract**

We describe a solution to a long standing puzzle about the difference in textures of quark CKM and lepton PMNS mixing matrices by deriving their common representation. We show how the difference in texture of the two matrices arises from assignment of lepton and quark pairs to different representation of a discrete two element symmetry group. The symmetry is absent in the Standard Model. It appears if, instead of Dirac spinors, one describes fermions in terms of bi-spinors and induces essentially unique textures: tri-bimaximal for lepton and $O(\lambda)$ of Wolfenstein parameterization for quark mixing.

## 1. Introduction

The origin of the form of the quark and lepton mixing matrices and the difference in their textures has been a long-standing puzzle in particle Physics. Its resolution remains elusive and it is one of the top three mysteries of the Standard Model [1, 2]. Explanations of the textures of mixing matrices usually employ introduction of extra dynamics using either extra gauge or discrete degrees of freedom or additional space-time dimensions. Recently a number of attempts have been made to derive a unified framework for lepto-quark mixing. Some recent work explores quark-lepton complementarity [3], TBM-Cabibbo mixing [4], Wolfenstein parameterization for lepton sector [5], the use of $SU(5)$ GUT vector fermions [6], and of "yukawaons" in a version of the flavon approach [7].

In this letter we derive a common representation for the quark CKM and the lepton PMNS mixing matrices without introduction of additional dynamics. The difference in textures of the two matrices appears as a result of assignment of lepton and quark pairs to different multiplets of a two element discrete symmetry. The symmetry is not present in the Standard Model (SM). It appears if, instead of the Dirac spinors, we use a bi-spinor[2] representation of fermions discovered by Ivanenko and Landau [8], who used it to provide an alternative solution to Dirac's solution of the electron gyromagnetic problem [9]. We show that the within the context of bi-spinor fermions, which are simply products of two Dirac spinors, in a formalism whose connection to differential geometry was elucidated in [10, 11], mixing matrix textures are unique. Thus we prove a conjecture made in [12, 13, 14] about the common form of mixing for lepto-quarks. Details of the derivations will be given in [15].

We will work within the context of SM4, the sequential extension of the SM to four generations, eventually reducing the number of generations to three in a consistent way. The resulting mixing matrices are non-unitary and exhibit very specific relations between matrix elements. We show that, despite strong constraints, there is no conflict of the form of bi-spinor induced mixing matrices with the experiment. Our results are unaffected by recent arguments, based on the tentative detection of 125 GeV Higgs [16], that the perturbative fourth generation is ruled out [17]. They apply to the observed mixing matrices, regardless of the existence or non-existence to the fourth generation.

---

[1] E-mail: jourjine@pks.mpg.de.

[2] In the literature bi-spinors are also referred to as Ivanenko-Landau-Kähler spinors or Dirac-Kähler spinors.

## 2. Lepto-Quark Mixing

We begin with presentation of the main result. This will be followed by its justification. We will use an explicit example of the quark sector of the SM4 with massive Dirac neutrinos and point out the modifications that must be made in the lepton sector as we go along. Eventually we will set the number of generations to three in a generally covariant way.

After spontaneous symmetry breakdown the non-interacting quark part of the SM4 Lagrangian is given by a sum of the kinetic and mass terms

$$\mathcal{L} = \mathcal{L}_k + \mathcal{L}_m, \tag{1}$$

$$\mathcal{L}_k = \overline{Q}_i^A (i\partial) Q_i^A + \overline{u}_R^A (i\partial) u_R^A + \overline{d}_R^A (i\partial) d_R^A, \tag{2}$$

$$\mathcal{L}_m = -\left(\overline{Q}_1^A \mathrm{M}_u^{AB} u_R^B + \overline{Q}_2^A \mathrm{M}_d^{AB} d_R^B + c.c.\right), \quad A = 1, \cdots, 4, \tag{3}$$

where $Q_i^A = \{u_L^A, d_L^A\}$ denotes the four generations of left-handed $SU(2)$ doublets, while the four generations of right-handed $u_R^A$, $d_R^A$ are $SU(2)$ singlets. We suppress the $SU(3)$ dependencies, since they play no role in the following. We now assign index $A = 1, \cdots, 4$ to the four generations of SM4 lepto-quarks by ($u^A = u_L^A + u_R^A$, etc)

$$\left(u^1, u^2, u^3, u^4\right) = (u, c, t', t), \qquad \left(d^1, d^2, d^3, d^4\right) = (d, s, b', b), \tag{4}$$

$$\left(v^1, v^2, v^3, v^4\right) = \left(\nu_e, \nu_\mu, \nu_{e_4}, \nu_\tau\right), \qquad \left(e^1, e^2, e^3, e^4\right) = (e, \mu, e_4, \tau). \tag{5}$$

This turns out to be the most obvious choice. The remaining choices will be analyzed in detail in [15]. For convenience this assignment has the conventional numbering of the fourth and the third generations switched. However, masses of $t', b'$ and $e_4, \nu_{e_4}$ are not assumed to be smaller then masses of $t, b$ and $\tau, \nu_\tau$, respectively.

In the SM4 the four $4 \times 4$ complex mass matrices $\mathrm{M}_{u,d}^{AB}$ and their lepton analogs $\mathrm{M}_{l,\nu}^{AB}$ are arbitrary. Their diagonalization generates mass spectrum of lepto-quarks and CKM/PMNS mixing matrices. We will now take each of $\mathrm{M}_{u,d}^{AB}$, $\mathrm{M}_{l,\nu}^{AB}$ to have a specific form

$$\mathrm{M} = B_1 \mathcal{M} B_2. \tag{6}$$

Factors $B_a \in U(2) \times U(2)$, $a = 1, 2$ are taken as $4 \times 4$ block-diagonal matrices with the upper-left blocks of which mix only $A = 1, 2$, while the lower-right blocks mix only $A = 3, 4$. They have the same generic form for quark and lepton sectors

$$B = \begin{pmatrix} U_1 & 0 \\ 0 & U_2 \end{pmatrix}, \qquad U_k = \begin{pmatrix} x_k & y_k \\ z_k & w_k \end{pmatrix} \in U(2). \tag{7}$$

Factor $\mathcal{M}$ is assigned according to

$$\mathcal{M} = diag\left(m^\nu{}_1, m^\nu{}_2, m^\nu{}_3, m^\nu{}_4\right) \qquad - \qquad \text{for neutrinos}, \tag{8}$$

$$\mathcal{M} = diag\left(m^l{}_1, m^l{}_3\right) \oplus m^l \hat{\mathcal{M}}^l{}_{24} \qquad - \qquad \text{for charged leptons, and} \tag{9}$$

$$\mathcal{M} = m^q{}_1 \hat{\mathcal{M}}^q{}_{13} \oplus m^q{}_2 \hat{\mathcal{M}}^q{}_{24}, \quad q = u, d, \quad - \quad \text{for quarks.} \tag{10}$$

Here the three $2 \times 2$ matrices $\hat{\mathcal{M}}^{l,q}{}_{kn} \in U(1,1)$ are such that $\hat{\mathcal{M}}^{l,q}{}_{13}$ mix $A = 1, 3$, while $\hat{\mathcal{M}}^q{}_{24}$ mixes $A = 2, 4$. Each $\hat{\mathcal{M}}^{l,q}{}_{kn}$ is parameterized by a real parameter $\lambda$ and is generically given by

$$\hat{\mathcal{M}} = \begin{pmatrix} \cosh \lambda & \sinh \lambda \\ \sinh \lambda & \cosh \lambda \end{pmatrix} = \frac{1}{\sqrt{2}} \begin{pmatrix} 1 & 1 \\ -1 & 1 \end{pmatrix} \begin{pmatrix} e^{-\lambda} & 0 \\ 0 & e^{+\lambda} \end{pmatrix} \frac{1}{\sqrt{2}} \begin{pmatrix} 1 & -1 \\ 1 & 1 \end{pmatrix}. \tag{11}$$

Combining the eigenvalues of $\hat{\mathcal{M}}^{l,q}{}_k$ in (11) with (8 -10) we obtain the mass spectrum of SM4 with the chosen mass matrices (6 - 11)

$$\left( m_{\nu_e} = m^\nu{}_1, m_{\nu_\mu} = m^\nu{}_2, m_{\nu_{e_4}} = m^\nu{}_3, m_{\nu_\tau} = m^\nu{}_4 \right) \quad - \quad \text{for neutrinos,}$$

$$\left( m_e = m^l{}_1, m_\mu = m^l{}_1 e^{-\lambda_1}, m_{e_4} = m^l{}_3, m_{\nu_\tau} = m^l{}_1 e^{+\lambda_1} \right) \quad - \quad \text{for charged leptons,}$$

$$\left( m_u = m^u{}_1 e^{-\lambda^u_1}, m_c = m^u{}_2 e^{-\lambda^u_2}, m_{t'} = m^u{}_1 e^{+\lambda^u_1}, m_t = m^u{}_2 e^{+\lambda^u_2} \right) \quad - \quad \text{for up quarks, and}$$

$$\left( m_d = m^d{}_1 e^{-\lambda^d_1}, m_s = m^d{}_2 e^{-\lambda^d_2}, m_{b'} = m^d{}_1 e^{+\lambda^d_1}, m_b = m^d{}_2 e^{+\lambda^d_2} \right) \quad - \quad \text{for down quarks.}$$

Since $\lambda$-parameters are arbitrary, this mass spectrum is unconstrained. It should be compared with the one derived in [13][3]. There a less general case with $m^q{}_1 = m^q{}_2$ for quark and lepton sectors was considered, which resulted in unobserved constraints on lepto-quark masses.

The particular form (6 - 11) of mass matrices for quarks and leptons results in the specific form of SM4 CKM/PMNS mixing matrices. These are obtained from field redefinitions needed to diagonalize (6). Quark mixing matrix is defined as $V = U_L D_L^+$, where $U_L, D_L$ are transformations for up/down quarks that transform the $u_L^A, d_L^A$ fields in (2, 3) into mass basis $u_{mL}^A = (U_L)^{AB} u^B{}_L$, $d_{mL}^A = (D_L)^{AB} d^B{}_L$. Lepton mixing matrix is defined analogously as $U = E_L N_L^+$, where $E_L, N_L$ enter mass basis transforms $e_{mL}^A = (E_L)^{AB} e^B{}_L$, $v_{mL}^A = (N_L)^{AB} v^B{}_L$.

We see from (6) that a diagonalizing field redefinition $W$ decomposes into two factors, $W = W_B U_P$, where $U_P$ removes one of the factors $B_a$ in (6), while $W_B$ diagonalizes $\mathcal{M}$. It follows from (8 - 10) that we must take

$$W_B = \begin{pmatrix} W_1 & 0 \\ 0 & W_2 \end{pmatrix}, \tag{12}$$

where $W_k = 1$, for neutrinos and $(e, e_4)$ pair, while, if we temporarily switch the $A = 2$, $A = 3$ indexes, for $(\mu, \tau)$ pair and all quark pairs

$$W_k = \frac{1}{\sqrt{2}} \begin{pmatrix} 1 & -1 \\ 1 & 1 \end{pmatrix}. \tag{13}$$

A the same time the second factor $U_P$ must be taken as

$$U_P = \begin{pmatrix} U_1 & 0 \\ 0 & U_2 \end{pmatrix}, \quad U_k = \begin{pmatrix} x_k & y_k \\ z_k & w_k \end{pmatrix} \in U(2), \tag{14}$$

where $|x_k| = |w_k| = \cos\varphi_k$, $|y_k| = |z_k| = \sin\varphi_k$, $\text{Arg } x_k + \text{Arg } w_k - \text{Arg } y_k - \text{Arg } z_k = \pi$. We conclude that (12-14) result in a generic lepto-quark mixing matrix $W^{SM4}$

$$W^{SM4} = W_B W_P (W_B)^+, \tag{15}$$

where $W_P \in U(2) \times U(2)$ is block-diagonal matrix defined as a product $U_u U_d^+$ or $U_e U_\nu^+$ in (14). Thus, we proved the conjecture made in [13, 14] about the common form of lepto-quark mixing. In terms of components of $U_k$, $\tilde{U}_k$ for quarks and, respectively, leptons we obtain

$$V_{CKM}^{SM4} = \frac{1}{2}\begin{pmatrix} x_1+x_2 & y_1+y_2 & x_1-x_2 & y_1-y_2 \\ z_1+z_2 & w_1+w_2 & z_1-z_2 & w_1-w_2 \\ x_1-x_2 & y_1-y_2 & x_1+x_2 & y_1+y_2 \\ z_1-z_2 & w_1-w_2 & z_1+z_2 & w_1+w_2 \end{pmatrix}, \quad U_{PMNS}^{SM4} = \begin{pmatrix} \tilde{x}_1 & \tilde{y}_1 & 0 & 0 \\ \tilde{z}_1/\sqrt{2} & \tilde{w}_1/\sqrt{2} & \tilde{z}_2/\sqrt{2} & \tilde{w}_2/\sqrt{2} \\ 0 & 0 & \tilde{x}_2 & \tilde{y}_2 \\ -\tilde{z}_1/\sqrt{2} & -\tilde{w}_1/\sqrt{2} & \tilde{z}_2/\sqrt{2} & \tilde{w}_2/\sqrt{2} \end{pmatrix}. \tag{16}$$

---

[3] Mass spectrum derivation there actually applies only to SM4 with a $U(2,2)$ mass term.

Unlike the general CKM matrix for the SM4 with six real parameters and three phases, $V_{CKM}^{SM4}$ has two real parameters and three phases [14]. It is easy to see that $U_{PMNS}^{SM4}$ has two real parameters and two phases.

The mixing matrices in (16) are now reduced to the three-generation case by removing from $A = 3$ row and column. This can be done in a generally covariant way by imposing constraints of the type $\det \Psi = 0$, where $\Psi$ is a quark/lepton bi-spinor field [18][4]. Briefly explained, $\det \Psi = 0$ implies after spinbein decomposition described in [12] that $\det \psi_\alpha^A = 0$, where $\psi_\alpha^A = u_\alpha^A$ or $\psi_\alpha^A = d_\alpha^A$ in (1-3) and similar for leptons. This means that using a linear field transformation we can set $u_\alpha^3 = d_\alpha^3 \equiv 0$ and eliminate the fourth generation from mixing matrices by dropping the $A = 3$ row and column. The removal $A = 3$ row and column with the help of $\det \Psi = 0$ constraint may appear *ad hoc* done by hand as an afterthought. This is justified to an extent. However, such removal of one generation of four is no more arbitrary then adding two generations by hand as it is done in the Standard Model. Important to realize is that to remove the physical one-particle states of the fourth generation the constraints must be imposed in the mass basis, that is diagonalization of (16) must be carried first.

We arrive at the final form of mixing matrices for the SM with massive Dirac neutrinos for the choice (6-11) of fermion mass matrices

$$V_{CKM} = \frac{1}{2}\begin{pmatrix} x_1+x_2 & y_1+y_2 & y_1-y_2 \\ z_1+z_2 & w_1+w_2 & w_1-w_2 \\ z_1-z_2 & w_1-w_2 & w_1+w_2 \end{pmatrix}, \quad U_{PMNS} = \frac{1}{2}\begin{pmatrix} \tilde{x}_1 & \tilde{y}_1 & 0 \\ \tilde{z}_1/\sqrt{2} & \tilde{w}_1/\sqrt{2} & \tilde{w}_2/\sqrt{2} \\ -\tilde{z}_1/\sqrt{2} & -\tilde{w}_1/\sqrt{2} & \tilde{w}_2/\sqrt{2} \end{pmatrix}. \tag{17}$$

Note that $V_{CKM}$ and $U_{PMNS}$ in (17) are not unitary. We also note that if $U_1 \approx U_2$ the values of $|V_{CKM}^{3A}|, |V_{CKM}^{A3}|$ for $A = 1, 2$ are suppressed. $U_{e3} \equiv U_{PMNS}^{e3}$ is suppressed as well: $U_{e3} = 0$ for any choice of $\tilde{U}_1, \tilde{U}_2$. $V_{CKM}, U_{PMNS}$ take an especially simple form if in (14) we take

---

[4] The second known method of generation reduction in bi-spinors [10] via projectors to left invariant ideals of Clifford algebra is not generally covariant.

$$U_1 = U_2, \quad \tilde{U}_1 = \begin{pmatrix} \sqrt{2/3} & \sqrt{1/3} \\ -\sqrt{1/3} & \sqrt{2/3} \end{pmatrix}, \quad \tilde{U}_2 = \begin{pmatrix} 1 & 0 \\ 0 & 1 \end{pmatrix}. \tag{18}$$

Then we obtain

$$V_{CKM}^0 = \begin{pmatrix} x & y & 0 \\ z & w & 0 \\ 0 & 0 & w \end{pmatrix}, \quad U_{PMNS}^0 = \begin{pmatrix} \sqrt{2/3} & \sqrt{1/3} & 0 \\ -\sqrt{1/6} & \sqrt{1/3} & \sqrt{1/2} \\ \sqrt{1/6} & -\sqrt{1/3} & \sqrt{1/2} \end{pmatrix}, \tag{19}$$

the former being the $O(\lambda)$, $\lambda = 0.2253 \pm 0.0007$, component of the Wolfenstein parameterization of the observed $V_{CKM}$, while the latter is the tri-bimaximal (TBM) mixing matrix. We will describe in [14] how the $U_{e3} \neq 0$ in (17) can arise from radiative corrections.

We now will explain briefly the choice of the mass matrices in (6 - 11). In SM4 it is arbitrary. However, it becomes severely restricted if we rewrite SM4 in terms of bi-spinors. In the four-generation bi-spinor extension of the SM the quark part of the Lagrangian (1 - 3) becomes

$$\mathcal{L}_k = \overline{\overline{Q}}_i^A (i\partial) Q_i^A + \overline{\overline{u}}_R^A (i\partial) u_R^A + \overline{\overline{d}}_R^A (i\partial) d_R^A, \tag{20}$$

$$\mathcal{L}_m = -\left( \overline{\overline{Q}}_1^A \mathbf{M}_u^{AB} u_R^B + \overline{\overline{Q}}_2^A \mathbf{M}_d^{AB} d_R^B + c.c. \right), \tag{21}$$

where $\overline{\overline{Q}}_i^A = \Gamma^{AB} \overline{Q}_i^B$, $\Gamma^{AB} = diag(1, 1, -1, -1)$, etc. Note that the last two indexes enter in the kinetic term with negative sign. As a result, it is not possible to extract from (20, 21) mass eigenstates by a diagonalization procedure. Instead one diagonalizes equations of motion. The end effect on mixing matrices is exactly the same as caused by the standard diagonalization procedure in SM4: the left- and the right- lepto-quarks undergo different linear transformations and thus generate mixing. Further, the requirement that masses are positive definite and coincide for left- and the right-handed fermions puts strong constraints on the form of mass matrices. It turns out that up to factors $B_a$, which appear as a result of bringing mass matrices into canonical Cartan form [13,14], there are essentially only five possible forms of $\mathcal{M}$ in (6) and hence of mass terms in (21). They correspond to multiplet assignments of lepto-quark pairs according to the value $s = 0, 1/2$ of a quantum number, called the scalar spin. One can show that, aside from arbitrary $B_a$ factors in (6), the possible tree-level mass terms are given by five cases for $\mathcal{M}$ (we refer for details to [15])

1. $(0,0)$: $\mathcal{M} = diag(m_A)$, $A = 1,\ldots,4$, $\quad m_A \neq m_B$,
2. $(0,1/2)$: $\mathcal{M} = diag(m_1, m_3) \oplus m \cdot U(1,1)$, $\quad m_A = m$, $\quad A = 2, 4$, $\quad m_1 \neq m_3$,
3. $(1/2,0)$: $\mathcal{M} = m \cdot U(1,1) \oplus diag(m_2, m_4)$, $\quad m_A = m$, $\quad A = 1, 3$, $\quad m_2 \neq m_4$,
4. $(1/2,1/2)$: $\mathcal{M} = m_1 \cdot U(1,1) \oplus m_2 \cdot U(1,1)$, $m_A = m_1$, $A = 1, 3$, $m_A = m_2$, $A = 2, 4$, $m_1 \neq m_2$,
5. $(1/2,1/2)$: $\mathcal{M} = m \cdot U(2,2)$, $\quad m_A = m$, $\quad A = 1,\ldots,4$,

where $(s = 0, 1/2, t = 0, 1/2)$ denotes assignment of scalar spin 0 or 1/2 to generation pairs $A = 1, 3$ or $A = 2, 4$, and $U(p, p)$ denotes an arbitrary $U(p, p)$ matrix.

The physical explanation for appearance of scalar spin is that bi-spinors transform in the representation of $SL(2, C)$ that is a direct product of $(0, 1/2) \oplus (1/2, 0)$ and its conjugate. As a result, when the Dirac spinor degrees of freedom are extracted from bi-spinors using the spinbein decomposition of a bi-spinor into Dirac spinor bilinears [12], the Lagrangian retains a remnant of the "right" spin quantum number introduced in [19]. After the extraction the remnant manifests itself in the form of discrete two-element symmetry of the Lagrangian (20, 21). The symmetry is not present in the SM or SM4. In bi-spinor SM or SM4 it can be used to classify mass eigenstates according to an additional quantum number.

Note now that our choice (8 - 11) selects case 1 for neutrinos, case 2 for leptons, and case 4 for quarks. Since case 3 is equivalent to case 2 and case 5 is the degenerate version of case 4, the assignment of mass matrices in bi-spinor SM is essentially unique, up to factors $B_a$. It is intriguing that in transition from quarks to leptons the mass symmetry reduction $U(1,1) \times U(1,1) \to U(1,1)$, parallels gauge symmetry reduction in the interaction term.

We now review the experimental data on $V_{CKM}$ and $U_{PMNS}$. First, let us see how well the characteristic bi-spinor SM relations $|V_{cs}| = |V_{tb}|$ and $|V_{ts}| = |V_{cb}|$ are satisfied. We should bear in mind that for comparison one must use direct measurements of mixing matrix elements or measurements where difference between $3 \times 3$ unitarity of the SM and $4 \times 4$ unitarity of SM4 is insignificant. From three experiments in neutrino scattering, semileptonic, and leptonic decays $|V_{cs}| = 0.94^{+0.32}_{-0.26} \pm 0.13$, $|V_{cs}| = 0.98 \pm 0.01_{exp} \pm 0.10_{theor}$, $|V_{cs}| = 1.030 \pm 0.038$, respectively. [20]. $|V_{tb}|$ has been measured directly giving $|V_{tb}| = 0.88 \pm 0.07$ [20]. More recently D0 reports $|V_{tb}| = 0.95 \pm 0.02$ [21]. In the SM we expect $|V_{tb}| = 0.999152^{+0.000030}_{-0.000045}$ [20].

As far as $|V_{ts}| = |V_{cb}|$ is concerned, from inclusive/exclusive semileptonic decays $|V_{cb}|_{inc} = (41.85 \pm 0.73) \cdot 10^{-3}$ and $|V_{cb}|_{excl} = (39.2 \pm 1.4_{exp} \pm 0.9_{theo}) \cdot 10^{-3}$ [22]. $|V_{ts}|$ cannot be determined directly. $|V_{ts}| = (38.7 \pm 2.1) \cdot 10^{-3}$ with partial use of SM unitarity in loop calculations and taking $|V_{tb}| = 1$ [20]. We note that experimentally the relations $|V_{cs}| = |V_{tb}|$ and $|V_{ts}| = |V_{cb}|$ are within a standard deviation in the experimental error bounds and at present one cannot distinguish the SM from bi-spinor SM on the basis of the relations.

Let us turn to $U_{PMNS}$. Until recently its TBM form (19) with $U_{e3} = 0$ was consistent with the experimental data. However, measurements of non-zero $U_{e3}$ were reported in [23, 24]. Unlike $|V^{3A}_{CKM}|, |V^{A3}_{CKM}| \le O(\lambda^2)$ for $A = 1, 2$, $|U_{e3}| = \sin\theta_{13} \approx 0.16$, $\theta_{13} = 9° \pm 1°$, is rather large and is $O(\lambda)$. Within our approach a possible explanation of the difference in magnitude comes from the form of quark and lepton mixing in (17). While for quarks the radiative corrections to $|V^{3A}_{CKM}| = |V^{A3}_{CKM}| = 0$ for $A = 1, 2$, contain subtractions of contributions of different generations, no such subtraction is present for leptons. Therefore, we expect that after quantum corrections $|U_{e3}| = O(\alpha)$, where $\alpha$ is the EW loop expansion parameter, while $|V^{3A}_{CKM}|, |V^{A3}_{CKM}| \le O(\alpha^2)$ for $A = 1, 2$.

As noted above, $V_{CKM}$ and $U_{PMNS}$ in (17) are not unitary. Presently $U_{PMNS}$ is not measured with sufficient accuracy to provide meaningful testing of the unitarity constraints. As for $V_{CKM}$, with one exception, violations of the SM unitarity due to SM4 unitarity are undetectable at present, because the discarded elements are suppressed by the same order of magnitude as the order of magnitude of the remaining suppressed elements. For the remaining unitarity constraints we expect violations on the order of $|V^{3A}_{CKM}|^2, |V^{A3}_{CKM}|^2$, $A = 1, 2$. This is less then $10^{-5}$ and currently undetectable, since the best measured unitarity constraint is $|V_{ud}|^2 + |V_{us}|^2 + |V_{ub}|^2 = 0.9999 \pm 0.0006$ [20]. The exception comes from the unitarity constraints that involve $V_{tb}$. However, the direct measurement of $V_{tb}$ comes with an insufficient precision to test this constraint either. In summary, either the SM or bi-spinor SM could be ruled out if the accuracy of the CKM elements measurements increases by one order of magnitude.